# Anharmonic Effect of Adiabatic Quantum Pumping

Wei-Yin Deng(邓伟胤), Ke-Ju Zhong(钟克菊), Rui Zhu(朱瑞), Wen-Ji Deng(邓文基)[†]

*Department of Physics, South China University of Technology, Guangzhou 510640, China*

*E-mail: [†]phwjdeng@scut.edu.cn*

Based on the scattering matrix approach, we systematically investigate the anharmonic effect of the pumped current in double-barrier structures with adiabatic time-modulation of two sinusoidal AC driven potential heights. The pumped current as a function of the phase difference between the two driven potentials looks like to be sinusoidal, but actually it contains sine functions of double and more phase difference. It is found that this kind of anharmonic effect of the pumped current is determined combinedly by the Berry curvature and parameter variation loop trajectory. Therefore small ratio of the driving amplitude and the static amplitude is not necessary for harmonic pattern in the pumped current to dominate for smooth Berry curvature on the surface within the parameter variation loop.

## 1 Introduction

In recent decades, there has been continued interest in the quantum pumping process [1-3]. Quantum pumping is a transport mechanism that DC charge or spin currents can flow under zero bias via a quantum system, in which some parameters are periodically modulated in time [4, 5]. It can be used as a standard for charge current because of the total pumped particle number equals integer times the Chern number of the scattering matrix within the parameter space in one cycle [6, 7]. The pumped current in various nanoscale systems is investigated since it was originally proposed by Thouless in 1983 [1], such as quantum dots [2, 8], mesoscopic one-dimensional wire [9-11], helical wire [12], mesoscopic rings [13], quasicrystals [14], bulk semiconductors [15,16], hybrid structures involving superconductors [17-19], and graphene-based systems [20-23]. Correspondingly, theoretical methods targeting quantum pumping problems are put forward by several groups [4, 5, 24-27].

The pumped current as a function of the phase difference between the two driven potentials looks like to be sinusoidal, but actually it contains sine functions of double and more phase difference. This non-sinusoidal current-phase relation is the anharmonic effect of pumped current. Double-barrier structures play a fundamental role in understanding the general physics of quantum pumping due to its direct physical picture and easy experimental realization [2, 20, 21, 28]. The non-sinusoidal behavior of adiabatic quantum pumping beyond linear response approximation in the double-barrier structure was studied by Zhu Rui [28]. By considering high-order expansion of the instant scattering matrix on the time, she found the anharmonic pumped current at large modulation amplitude [2, 24]. However, in this approach, expansion into even higher orders requires extremely tedious algebra and a correspondence between small amplitude expansion and the Berry curvature surface integral is missed. The Berry curvature concept was proposed by J. E. Avron et al. [26], they related the pumped charge to Berry's phase and the corresponding Brouwer pumping formula to curvature.

Partially for these reasons, in this work, we investigate the general anharmonic effect of an adiabatic quantum pumping. We take the double barrier structure as a showcase and the consideration can be transplanted into other configurations directly. Compared to the Ref. [28], we use two methods to analysis the anharmonic effect of pumped current, which are the Berry curvature and instant scattering matrix expansion from Brouwer pumping formula, and obtain a correspondence between small amplitude expansion and the Berry curvature surface integral.

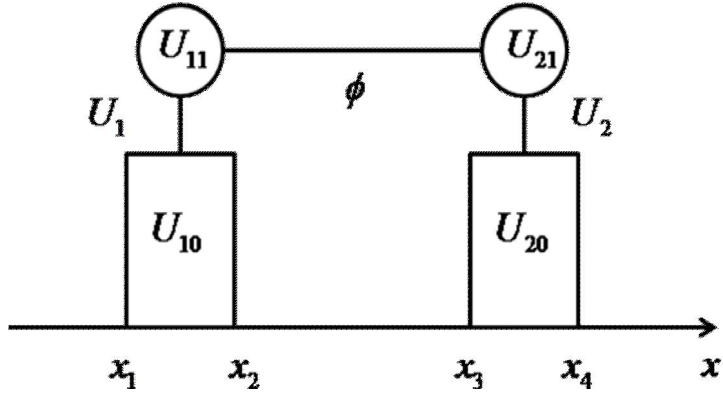

Fig. 1. Schematics of the quantum pumping: a nanowire modulated by two AC potentials. $\phi = \phi_1 - \phi_2$ is the phase difference between the two driven potentials.

## 2 Model and formulations

We consider a nanowire modulated by two AC potentials (see Fig. 1). The double-barrier model along the $x$ direction is used, in which the left and right parts are two semi-infinite leads, and they connect with the double-barrier structure. To generate a dc current at zero bias, the two time-modulated potential barriers are in the following form

$$U(x,\tau) = \begin{cases} U_1(\tau) = U_{10} + 2U_{11}\cos(\omega\tau + \phi_1), & x_1 \leq x \leq x_2, \\ 0, & \text{others}, \\ U_2(\tau) = U_{20} + 2U_{21}\cos(\omega\tau + \phi_2), & x_3 \leq x \leq x_4, \end{cases} \quad (1)$$

where $U_{j0}$ and $2U_{j1}$ ($j=1,2$) denote the static potential and the driving amplitude, respectively; $\tau$ is the time, $\omega$ is the frequency and $\phi_j$ is the initial phase of the driver. The Hamiltonian follows directly as

$$\hat{H} = -\frac{\hbar^2}{2m}\frac{\partial^2}{\partial x^2} + U(x,\tau), \tag{2}$$

where $m$ is the effective mass of electron.

In the adiabatic situation, the barrier height varies so slowly that the physical process in one period can be considered as the accumulated effect of the instant stationary processes. By defining the dimensionless variables $\hat{H} \to \hat{H}/E_F$ and $x \to k_F x$ with $k_F = \sqrt{2mE_F}/\hbar$, the wave functions in different regions can be formulated as

$$\psi(x) = \begin{cases} a_L e^{i(x-x_1)} + b_L e^{-i(x-x_1)}, & x \le x_1, \\ A_2 e^{\kappa_1(x-x_1)} + B_2 e^{-\kappa_1(x-x_1)}, & x_1 \le x \le x_2, \\ A_3 e^{i(x-x_2)} + B_3 e^{-i(x-x_2)}, & x_2 \le x \le x_3, \\ A_4 e^{\kappa_2(x-x_3)} + B_4 e^{-\kappa_2(x-x_3)}, & x_3 \le x \le x_4, \\ a_R e^{-i(x-x_4)} + b_R e^{i(x-x_4)}, & x \ge x_4. \end{cases} \tag{3}$$

Here $\kappa_1$ and $\kappa_2$ are dimensionless wave vectors, which can be obtained from the dispersion relations $U_1 - \kappa_1^2 = U_2 - \kappa_2^2 = 1$. $A_i, B_i\ (i=2,3,4)$ are the wave function amplitudes in the middle region. $a_\alpha$ and $b_\alpha$ $(\alpha = L, R)$ are the incident and outgoing amplitudes respectively. The scattering matrix can be given from

$$\begin{pmatrix} b_L \\ b_R \end{pmatrix} = S \begin{pmatrix} a_L \\ a_R \end{pmatrix}, \tag{4}$$

with $S = [r, t'; t, r']$. The elements of the scattering matrix can be obtained from the continuity relation as

$$r = -2i\frac{\sinh(\kappa_2 b)(\kappa_2 + \kappa_2^{-1})(D_1^* e^{\kappa_1 b} + D_1 e^{-\kappa_1 b})e^{ia} + \sinh(\kappa_1 b)(\kappa_1 + \kappa_1^{-1})(D_2 e^{\kappa_2 b} + D_2^* e^{-\kappa_2 b})e^{-ia}}{4\sinh(\kappa_1 b)\sinh(\kappa_2 b)(\kappa_1 + \kappa_1^{-1})(\kappa_2 + \kappa_2^{-1})e^{ia} + (D_1 e^{\kappa_1 b} + D_1^* e^{-\kappa_1 b})(D_2 e^{\kappa_2 b} + D_2^* e^{-\kappa_2 b})e^{-ia}},$$

$$t = \frac{16}{4\sinh(\kappa_1 b)\sinh(\kappa_2 b)(\kappa_1 + \kappa_1^{-1})(\kappa_2 + \kappa_2^{-1})e^{ia} + (D_1 e^{\kappa_1 b} + D_1^* e^{-\kappa_1 b})(D_2 e^{\kappa_2 b} + D_2^* e^{-\kappa_2 b})e^{-ia}}.$$

(5)

where $D_1 = 2 + i\kappa_1 - i\kappa_1^{-1}$, $D_2 = 2 + i\kappa_2 - i\kappa_2^{-1}$, $a = x_3 - x_2$, $b = x_2 - x_1 = x_4 - x_3$, $t' = t$, $r' = -tr^*/t^*$. And the conservation of probability holds as $|r|^2 + |t|^2 = 1$.

In the adiabatic regime, the pumped current from the left reservoir to the right could be expressed in terms of the instant scattering matrix as follows [5, 24]:

$$I_L = \frac{e\omega}{4\pi^2 i} \int_0^{\tau_0} d\tau \left(\frac{\partial S}{\partial t} S^+\right)_{LL}. \tag{6}$$

where $\tau_0$ is one cycle time. We can change the integral variable:

$$I_L = \frac{e\omega}{4\pi^2 i} \oint_A \left[\frac{\partial S}{\partial U_1} S^+ dU_1 + \frac{\partial S}{\partial U_2} S^+ dU_2\right]_{LL}, \tag{7}$$

where the integral is around the cyclic loop $A$ of the variation of $(U_1, U_2)$. It can be transformed to the surface integral form by the Green's theorem:

$$I_L = \frac{e\omega}{4\pi^2} \iint_C J(U_1, U_2) dU_1 dU_2, \tag{8}$$

here the Berry curvature [3, 26] of the scattering matrix is

$$J(U_1, U_2) = i\left(\frac{\partial S}{\partial U_1}\frac{\partial S^+}{\partial U_2} - \frac{\partial S}{\partial U_2}\frac{\partial S^+}{\partial U_1}\right)_{LL}. \tag{9}$$

The surface integral form of pumped current is easy to analyse the harmonic effect because the elliptical area $C$ in $(U_1, U_2)$ parameter space is proportional to $\sin(\phi_1 - \phi_2)$ under the harmonic driving barriers as Eq. (1). For special values $0$, $\pi$ and $2\pi$ of phase difference $\phi$, the pumped current is zero because the elliptical area of integration keeps being zero.

Another way to analyze the anharmonic effect of the pumped current is the Taylor's expansion of the instant scattering matrix in the parameter space at equilibrium parameter values.

$$\begin{aligned}S = S_0 &+ \sum_{i=1}^{2} \left.\frac{\partial S}{\partial U_i}\right|_0 \Delta U_i + \sum_{i=1}^{2}\sum_{j=1}^{2} \frac{1}{2} \left.\frac{\partial^2 S}{\partial U_i \partial U_j}\right|_0 \Delta U_i \Delta U_j \\ &+ \sum_{i=1}^{2}\sum_{j=1}^{2}\sum_{k=1}^{2} \frac{1}{6} \left.\frac{\partial^3 S}{\partial U_i \partial U_j \partial U_k}\right|_0 \Delta U_i \Delta U_j \Delta U_k + \cdots.\end{aligned} \tag{10}$$

Only if the order $n+1$ is smaller than the order $n$ of the series in Eq. (10), the Taylor expansion is valid. More specifically, the Berry curvature in the parameter space should change slowly and smaller modulation amplitude of the driving signals is better. Substituting Eq. (10) into Eq. (6), we can obtain the pumped current of all orders as

$$I_L = I_{0L} + I_{1L} + I_{2L} + I_{3L} + \cdots. \tag{11}$$

The zero and first order pumped current $I_{0L} = 0$, $I_{1L} = 0$. The second order pumped current is

$$I_{2L} = -\frac{e\omega}{\pi i}\left[\left(\frac{\partial S}{\partial U_1}\frac{\partial S^+}{\partial U_2} - \frac{\partial S}{\partial U_2}\frac{\partial S^+}{\partial U_1}\right)\bigg|_0\right]_{LL} U_{11}U_{21}\sin(\phi_1 - \phi_2). \tag{12}$$

This is the pumped current in the weak pumping limit [24]. The second approximation pumped current is sinusoidal behavior so the anharmonic effect of the current should be found in much higher order. The third order current $I_{3L} = 0$ and the fourth order is

$$I_{4L} = -\frac{e\omega}{2\pi i}\left\{\left[\begin{array}{l}U_{11}^3 U_{21}\sin(\phi_1 - \phi_2)\left(2\frac{\partial^2 S}{\partial U_1^2}\frac{\partial^2 S^+}{\partial U_1 \partial U_2} + \frac{\partial S}{\partial U_1}\frac{\partial^3 S^+}{\partial U_1^2 \partial U_2} + \frac{\partial^3 S}{\partial U_1^3}\frac{\partial S^+}{\partial U_2} - H.C.\right)\\ -U_{11}U_{21}^3\sin(\phi_1 - \phi_2)\left(2\frac{\partial^2 S}{\partial U_2^2}\frac{\partial^2 S^+}{\partial U_1 \partial U_2} + \frac{\partial S}{\partial U_1}\frac{\partial^3 S^+}{\partial U_2^3} + \frac{\partial^3 S}{\partial U_1 \partial U_2^2}\frac{\partial S^+}{\partial U_2} - H.C.\right)\\ +U_{11}^2 U_{21}^2\sin[2(\phi_1 - \phi_2)]\left(\frac{\partial^2 S}{\partial U_1^2}\frac{\partial^2 S^+}{\partial U_2^2} + \frac{\partial S}{\partial U_1}\frac{\partial^3 S^+}{\partial U_1 \partial U_2^2} + \frac{\partial^3 S}{\partial U_1^2 \partial U_2}\frac{\partial S^+}{\partial U_2} - H.C.\right)\end{array}\right]\bigg|_0\right\}_{LL}.$$

(13)

The fourth order pumped current is sine functions of phase difference and double phase difference. The part of sine function of double phase difference is the anharmonic effect. It can be seen that the anharmonic effect of the current also occurs in higher orders. In addition,

If the ratio of $\omega_1/\omega_2 = n_1/n_2$ is a rational number, then the whole system is still periodic with a new frequency $\omega$ depending on $\omega_1$ and $\omega_2$. These cases can be easily handled within the present theory of quantum pumping, and the only distinction is that the integral path is more complex Lissajous figure instead of an ellipse, the simple Lissajous figure with $\omega_2 = \omega_1$. At this time the harmonic behavior would disappear.

## 3  Results and discussions

Below, we consider a particular quantum pumping system of a double barrier structure and see how the system parameters shapes the Berry curvature and hence the pumped current. By tuning the Fermi energy $E_F$, the width $b$ of the barrier region, the width $a$ of the middle region and the barrier strength $U_j$, we carry out a series of numerical simulations and results are shown in Figs. 2 ~ 5. For a quantum wire fabricated on a GaAs two-dimensional electron gas, the effective mass $m = 0.067 m_e$. We set the width $a = 1.40 \left( L_a = 4.31 \text{nm} \right)$ and the Fermi energy to be $60 \text{meV}$ according to the resonant level within the double-barrier structure, the static magnitude of the two gate potentials $U_{10} = U_{20} = U_0$, the AC driving amplitude of the modulations equal $U_{11} = U_{21} = U_\omega$ and the phase difference $\phi = \phi_1 - \phi_2$.

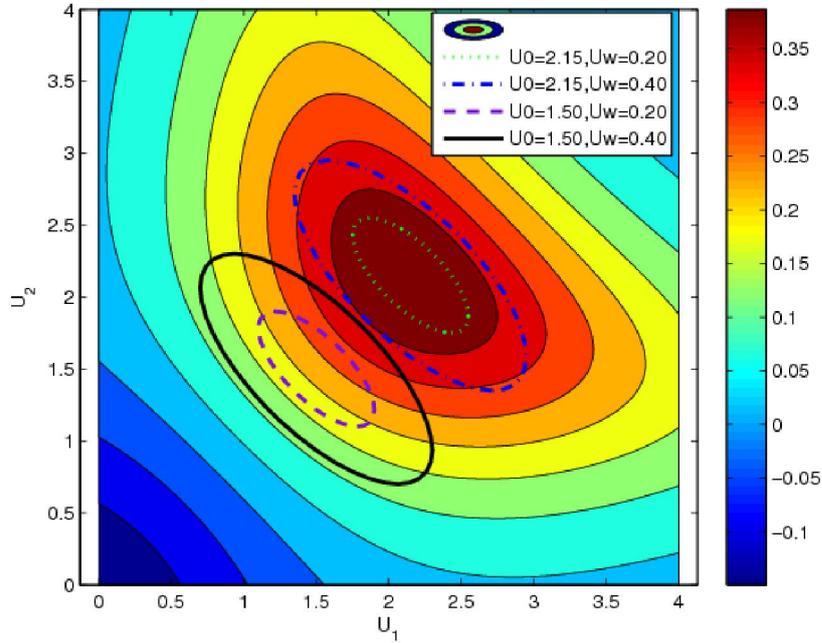

Fig. 2. The Berry curvature $J$ as functions of the two parameters $U_1$ and $U_2$. The width of the barrier region $b = 0.65 \left( L_b = 2 \text{nm} \right)$.

In Fig. 2, we plot the Berry curvature $J(U_1, U_2)$ in the parameter space according to Eqs.

(5) and (9) with $b = 0.65 (L_b = 2\text{nm})$. The green dotted line, blue dash-dotted line and purple dashed line correspond to the loop with area at the phase difference $\phi = 0.75\pi$ with $(U_0, U_\omega) = (2.15, 0.20)$, $(2.15, 0.40)$ and $(1.50, 0.20)$, respectively. From Eq. (8) we know that the pumped current is the integral of the Berry curvature inside the area $C$. If the Berry curvature varies so slowly within $C$ that it can be approximated by a constant $J_0$, the pumped current

$$I_L \approx \frac{e\omega}{4\pi^2} J_0 \iint_C dU_1 dU_2 \sim \frac{e\omega}{4\pi^2} J_0 \sin\phi. \tag{14}$$

That is the harmonic effect of the pumping current generated by single-photon processes [24]. The essential of the harmonic effect of the pumped current is that the Berry curvature function is a constant in the $(U_1, U_2)$ parameter space. Just as the $(U_0, U_\omega) = (2.15, 0.20)$, the pumped current as a function of the phase difference is sinusoidal, which is shown in Fig. 3 (a). So the anharmonic effect of pumped current is related to the derivative of $S$ and $U_j$. When the AC driving amplitude of the modulations $U_\omega$ is increased, the elliptical area $C$ would be increased. Such as the $(U_0, U_\omega) = (2.15, 0.40)$, the pumped current demonstrates weak anharmonic effect [see Fig. 3 (b)], the second order Fourier component is not zero. At the same static potential $U_0$, the anharmonic effect of the pumped current is always more prominent for larger AC driving amplitude $U_\omega$. In the red dashed loop of Fig. 2, we change the static potential $U_0$ with $(U_0, U_\omega) = (1.50, 0.20)$. It is found that stronger anharmonic effect occurs [see Fig. 3 (c)] compared to the previous two loops, the second order Fourier component is large. So small $U_\omega / U_0$ is not a necessary requirement to generate harmonic pumped current. It depends on the smoothness of the Berry curvature in the parameter space around $U_0$.

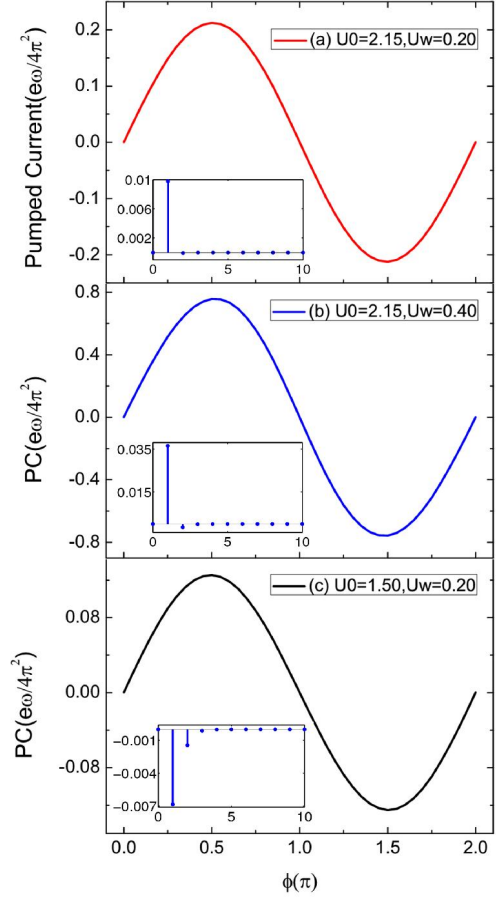

Fig. 3. Pumped current (PC) $I_L$ versus the driving phase difference $\phi$ for different cyclic loops $(U_0, U_\omega)$. The width of the barrier region $b = 0.65 (L_b = 2\text{nm})$. Insets are Fourier components $a_n$ of the curves, the horizontal axis shows $n$, where $a_n = \pi^{-1} \int_0^{2\pi} I_L(\phi) \sin n\phi \, d\phi$.

$b_n = \pi^{-1} \int_0^{2\pi} I_L(\phi) \cos n\phi \, d\phi$ are another part of Fourier components, which are zero in the numerical calculations.

In order to see the anharmonic effect of the pumped current more clearly, we plot the pumped current as a function of the phase difference with $(U_0, U_\omega) = (1.50, 0.40)$ in Fig. 4. On the one hand, we use the Taylor's expansion to analyze the anharmonic effect. In Fig. 4 the red solid line, blue dashed line and black dotted line correspond to the total, the fourth-order approximated and the second-order approximated currents, respectively. The anharmonic effect is obvious from the Fourier spectrum, the second order Fourier component is large and the third one is not zero. The first two harmonics dominates in the total current. Higher orders of the Taylor's expansion diminish as high-order derivatives of the scattering matrix diminish exponentially. On the other hand, we can use the Berry curvature to do qualitative analysis. $J$ changes abruptly inside the area $C$ with $(U_0, U_\omega) = (1.50, 0.40)$ at the phase difference $\phi = 0.75\pi$, so strong anharmonic effect is obvious.

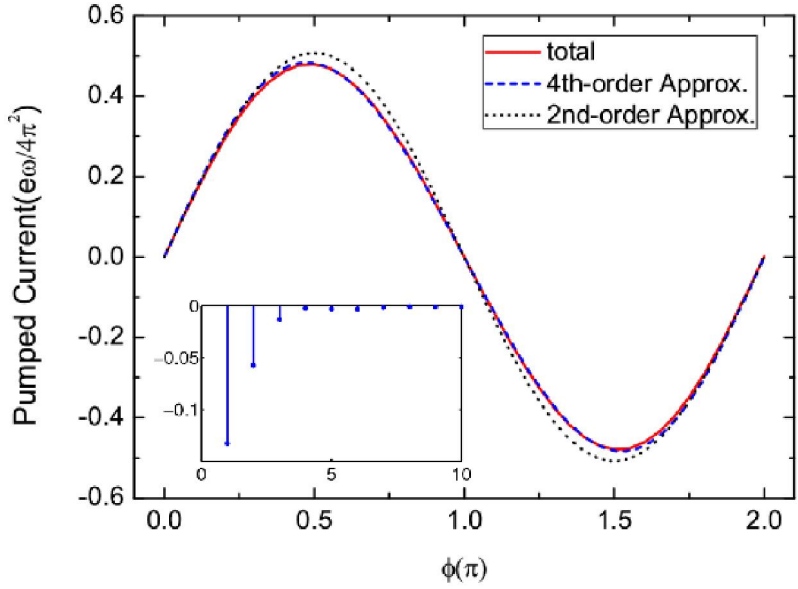

Fig. 4. Pumped current $I_L$ versus the driving phase difference $\phi$ for different accuracies. The chosen parameters are $U_0 = 1.50$, $U_\omega = 0.40$ and $b = 0.65 (L_b = 2\text{nm})$. Insets are Fourier components of the curves just as Fig. 3.

In Fig. 5, we plot the pumped current as function of the phase difference with wider barrier region $b = 1.30 (L_b = 4\text{nm})$ and $(U_0, U_\omega) = (2.15, 0.40)$. Compared to the Fig. 3 (b), the second Fourier component of the total PC curve is more obvious. The width of the barrier region $b$ makes an important role in the anharmonic effect of pumped current. This is because $b$ is related to the scattering matrix which is related to the Berry curvature. What's more, we can find that in this configuration the fourth-order approximation is not corresponded to the total one. At this time the higher order pumped current is more important than the lower one, such as the fourth order PC is bigger than the second order PC. We can't use the Taylor expansion to analyze the anharmonic effect at this situation.

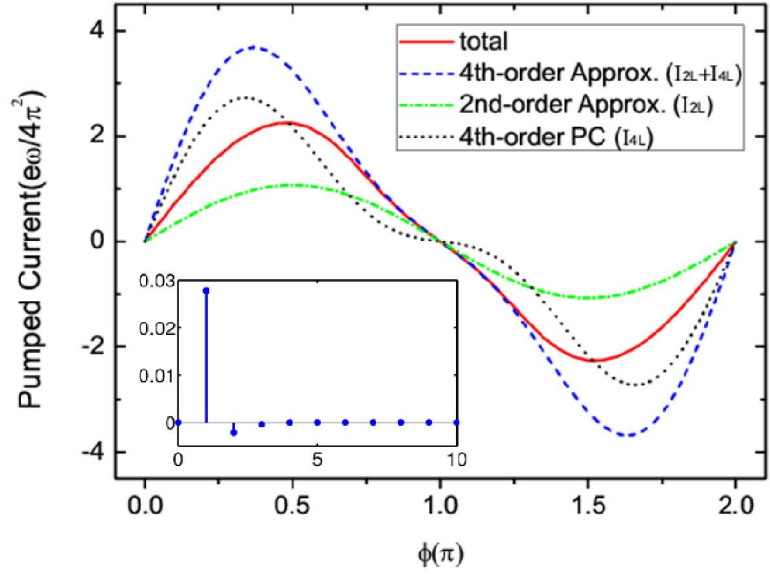

Fig. 5. Pumped current (PC) $I_L$ versus the phase difference $\phi$ for different approximation. The chosen parameters are $U_0 = 2.15$, $U_\omega = 0.40$ and $b = 1.30\,(L_b = 4\text{nm})$. Insets are Fourier components of the curves just as Fig. 3.

## 4  Conclusion

In summary, we have investigated the anharmonic effect of pumped current in the double-barrier structure with the adiabatic time-modulation by two AC driven potential barriers. Based on the scattering approach with the analytic expressions of the reflection and transmission, the pumped current as a function of difference phase for different parameters were calculated, such as the static magnitude of the two gate potentials $U_0$, the AC driving amplitude of the modulations $U_\omega$ and the width of the barrier region $L_b$; and the relationship between the structure parameters and anharmonic effect of the pumped current were discussed by the Berry curvature surface integral and instant scattering matrix expansion. It is found that this kind of anharmonic effect is not only related to $U_\omega$ and the derivative of the scattering matrix to the potential, but also related to $L_b$. Small ratio of $U_\omega/U_0$ is not necessary for harmonic pumped current because it depends on the smoothness of the Berry curvature in the parameter space around $U_0$.

Acknowledgements

This project was supported by the National Natural Science Foundation of China (No 11004063) and the Fundamental Research Funds for the Central Universities, SCUT (No 2012ZZ0076).